\documentclass[a4paper,fleqn,usenatbib,useAMS]{mnras}

\usepackage{graphicx}	
\usepackage{amsmath}	
\usepackage{amssymb}	
\usepackage{multicol}        
\usepackage{bm}		
\usepackage{pdflscape}	

\usepackage{color}
\newcommand{\Msun}{{\rm ~M}_{\odot}}

\newcommand{\Zsun}{{\rm ~Z}_{\odot}}
\newcommand{\gpy}{{\rm ~Gpc}^{-3} {\rm ~yr}^{-1}}

\defcitealias{Chruslinska18}{C18}
\defcitealias{Belczynski16N}{B16}

\usepackage[T1]{fontenc}
\usepackage{ae,aecompl}

\usepackage{txfonts}

\title[SFR(Z,z) and the properties of merging DCO]{The influence of the distribution of cosmic star formation 
at different metallicities on the properties of merging double compact objects}

\author[M. Chruslinska et al.]
{Martyna Chruslinska,$^{1}$\thanks{E-mail: m.chruslinska@astro.ru.nl}
Gijs Nelemans$^{1,2}$, Krzysztof Belczynski$^{3}$
\\
$^{1}$Department of Astrophysics/IMAPP, Radboud University, P O Box 9010, NL-6500 GL Nijmegen, The Netherlands\\
$^{2}$Institute for Astronomy, KU Leuven, Celestijnenlaan 200D, 3001 Leuven, Belgium\\
$^{3}$Nicolaus Copernicus Astronomical Center, Polish Academy of Sciences, Bartycka 18, 00-716 Warsaw, Poland}
\date{Last updated 2015 May 22; in original form 2013 September 5}

\pubyear{2018}
\begin{document}
\label{firstpage}
\pagerange{\pageref{firstpage}--\pageref{lastpage}}
\maketitle

\begin{abstract}
Binaries that merge within the local Universe originate from progenitor systems that formed at different times
and in various environments.
The efficiency of formation of double compact objects is highly sensitive to metallicity of the star formation. 
Therefore, to confront the theoretical predictions with observational limits resulting from gravitational waves
observations one has to account for the formation and evolution of progenitor stars in the chemically evolving Universe.
In particular, this requires knowledge of the distribution of cosmic star formation rate at different metallicities and
times, probed by redshift (SFR(Z,z)). 
We investigate the effect of the assumed SFR(Z,z) on the properties of merging double compact objects, 
in particular on their merger rate densities.
Using a set of binary evolution models from Chruslinska et al. (2018) we demonstrate that the reported tension between the 
merger rates of different types of double compact objects and current observational limits
in some cases can be resolved 
if a SFR(Z,z) closer to that expected based on observations of local star-forming galaxies is used,
without the need for changing the assumptions about the evolution of progenitor stars of different masses. 
This highlights the importance of finding tighter constraints on SFR(Z,z) and understanding the associated uncertainties.
\end{abstract}

\begin{keywords}
 stars: binaries - stars: black holes - stars: neutron - gravitational waves
\end{keywords}



\section{Introduction}

Metallicity is the second most important property, just after mass, determining the stellar evolution.
It affects, among others, stellar winds and radii, also impacting the evolution of stars in binaries 
and the outcome of their evolution \citep[e.g.][]{Maeder92,Hurley00,Baraffe01,Vink01,Belczynski10Mmax}.
In particular, the number of close double compact binaries of certain type created per unit of mass formed in stars is known to vary
depending on the composition of progenitor stars, the effect being especially significant for double black holes 
\citep[e.g.][]{Belczynski10,Dominik12,EldridgeStanway16,Stevenson17,Klencki18,Giacobbo18}.
Such binaries are the main astrophysical source of gravitational waves that are detected with the currently operating network 
of ground-based gravitational wave detectors \citep{Abbott16_limits}.
Using information obtained with detections of gravitational waves from their mergers 
(e.g. limits on their merger rate density) one can gain insight on the evolution of progenitors of compact binaries.
This can be done by confronting theoretically calculated merger rate densities,
strongly dependent on the assumptions made in order to describe poorly understood stages of binary evolution 
(e.g. common envelope evolution, core-collapse events and related natal kick velocities) with observational limits
\citep[see e.g.][for recent results]
{Chruslinska18,Giacobbo18,Barrett18} 
.\\
However, since double compact objects (DCO) can form with different parameters (masses, separations, eccentricities), 
they need different amount of time to merge due to gravitational wave radiation \citep[e.g.][]{Peters64}.
As a consequence, binaries formed at different times, in different environments and hence with different metal content
all contribute to the merger rate we measure locally.
The number of merging binaries depends on the amount of star formation happening throughout the
cosmic time (probed by redshift) but also on the distribution of the star formation rate
across different metallicities (SFR(Z,z); since the evolution is metallicity dependent).
To estimate the merger rate density it is necessary to assume a certain history of star formation and 
chemical evolution of the Universe, which adds another layer of uncertainty to those calculations.\\
Alterations in the assumed SFR(Z,z) also change the properties of the locally merging population of DCO.
Since the dependence of the formation efficiency of merging double compact objects 
on metallicity is different for different types of binaries, any change in SFR(Z,z) particularly affects the ratio of their merger rates.
Moreover, as the stellar wind mass loss is a function of metallicity, changing SFR(Z,z) will
have a significant effect on the distribution of masses of the locally merging double black holes.
\\
Recently, \citet{Chruslinska18} (hereafter \citetalias{Chruslinska18})
demonstrated that the local double neutron star (NSNS) merger rate densities typically
fall significantly below the current lower limit implied by gravitational wave observations \citep{GW170817}. 
Within a set of 21 models calculated with the StarTrack population synthesis code \citep{Belczynski02,Belczynski08}
they identify three, requiring quite extreme assumptions about the evolution
of progenitor stars that lead to NSNS merger rates consistent with this limit.
However, the associated double black hole merger rate densities calculated for those models
exceed the upper limit on their merger frequency set by LIGO/Virgo observations \citep{GW170104}.
We argue that the assumed distribution of the cosmic star formation at different metallicities and redshifts
used in this study significantly over predicts the amount of star formation happening at low metallicities.
We use their models as an example to demonstrate the consequences of
different assumptions on SFR(Z,z) for the properties of merging DCOs.
We show that in two out of three cases the reported discrepancy 
may be resolved if a different SFR(Z,z) with higher metallicity of the star formation is used.
\\
Throughout the paper we adopt a standard flat cosmology with the following cosmological parameters:
$H_{0}$ = 70 $km \ s^{-1} \ Mpc^{-1}$, $\Omega_{M}$=0.3, $\Omega_{\Lambda}$=0.7 and $\Omega_{k}$=0.

\section{Merger rate density and metallicity}\label{sec: rloc and Z}
The rate of DCO mergers is strictly connected to the pace of the cosmic star formation (SFR(z)).
The higher the SFR(z), the more DCO mergers. Those mergers occur with a certain delay in time,
which is needed to complete the evolution of stars  
to the point where two compact objects coalesce.
In general, this time depends on the binary parameters (and as such is metallicity dependent).
The distribution of DCO delay times $t_{del}$ is typically strongly peaked at short times ($\sim$100 Myr) 
and falls off as $\sim t_{del}^{-\alpha}$, with $\alpha$ being of the order of unity.
Knowing this distribution, one can calculate what fraction ($\rm f_{loc}^{mr}$) of merging systems formed at redshift
$z$ with metallicity $Z$ merges in the local Universe (at redshift $z\leq z_{loc}$ or
equivalently within $\delta t_{loc} = t_{0} - t(z_{loc})$, where $t(z=0)=t_{0}$ is the Hubble time).
However, since the efficiency of formation of merging DCOs ($\rm \chi_{DCO;i}$)\footnote{$\rm \chi_{DCO;i}$ 
is defined as a number of merging double compact objects of certain type created per unit of mass formed in stars} 
is a function of metallicity, it is not enough to know the absolute star formation rate (SFR(z)), but rather its distribution
at different metallicities SFR(Z,z) or a fraction of SFR(Z) that at each $z$ happens at 
a certain metallicity ($\rm f_{sfr}(z,Z)$).\\
Thus, the local merger rate density of DCOs of certain type 
(double neutron stars - NSNS, double black holes - BHBH, neutron star - black hole binaries BHNS/NSBH)
can be expressed as:
\begin{equation}\label{eq: Rloc}
\begin{aligned}
 \rm R^{loc}_{\rm DCO;i} = \frac{1}{\delta t_{loc}} \sum_{z} \sum_{Z} ( \ \chi_{DCO;i}(Z) \ f_{sfr}(z,Z) \ \frac{SFR(z)}{\delta V} \ \times \\
 \left[t(z+\delta z) - t(z)\right] \ f_{loc}^{mr}(z,Z) \ )
\end{aligned}
 \end{equation}
where the sum runs over redshifts (z) and metallicities (Z) at which the progenitor stars form and $\delta V$ 
is the comoving volume element.
Note that $\rm \chi_{DCO;i}$ is model\footnote{in population synthesis studies a model is defined by the choice of a particular set 
of assumptions (parameters) used to describe the evolution of binaries, e.g. conservativeness of the mass transfer, 
distribution describing the magnitude of NS and BH natal kicks} dependent. 
Modifications of the assumptions made to describe evolution of DCO progenitors result in changes in
 $\rm \chi_{DCO;i}$ and hence in the estimated merger rates, e.g. if more neutron stars are allowed to form with relatively 
 small natal kick velocities, 
 the formation efficiency of merging double neutron stars generally increases. 
 However, certain modifications can boost or decrease $\rm \chi_{DCO;i}$ only in specific
 metallicity range \citep[e.g. assuming fully-conservative mass transfer as discussed in][\citetalias{Chruslinska18}
 (models V12 and J5 respectively) affects mostly the number of merging NSNS formed at high metallicity]{Dominik12}.\\
 For a given model, $\rm \chi_{DCO}(Z)$ also depends on the choice of distributions used to describe the initial 
 parameters of binaries \citep{deMinkBelczynski15}, although the change is minor, unless the high mass tail of the initial mass function 
 is allowed to vary with metallicity \citep[see fig. 6 in][]{Klencki18}.
 Despite this sensitivity of $\rm \chi_{DCO}(Z)$ on the model assumptions, certain characteristics seem robust
 (see \citealt{Giacobbo18} and sec. 4.2 in \citealt{Klencki18}).
 For instance, BHBH form much more efficiently at low metallicities than at high $Z$ and $\rm \chi_{BHBH}(Z)$ 
 reveals a sharp decrease (a factor of $\gtrsim$10)
 at $Z$ approaching solar values. $\rm \chi_{NSNS}(Z)$ usually shows much smaller variation with metallicity 
 and increases slightly towards higher $Z$ \footnote{but see models J1$B$, J7$B$ and J5$B$ in \citetalias{Chruslinska18},
 where $\rm \chi_{NSNS}(Z)$ decreases at high $Z$ - this can be seen by comparing the numbers in column 3 from their table 2, 
 However, note that those models significantly underpredict the Galactic merger rates}.
 Generally, $\rm \chi(Z)$ evolves differently for different types of DCOs, hence any change in $\rm f_{sfr}(z,Z)$ 
 would affect the ratios of merger rates of binaries of different type.\\
 Different approaches have been taken to determine $\rm f_{sfr}(z,Z)$ used to calculate merger rate densities.
 One way is to extract this information from cosmological simulations \citep[e.g.][]{Mapelli17,Schneider17}, 
 the other is to use the available observations and/or complement observational results with theoretical inferences
 \citep[][Chruslinska et al. in prep.]{Dominik13,Belczynski16N,Eldridge18}.

 All methods have their shortcomings. Cosmological simulations do not fully reproduce all of the observational relations 
 (e.g. mass - metallicity relation) and are resolution-limited. 
 Observations on the other hand are subject to biases and provide complete information only 
 in very limited ranges of redshifts and luminosities of the objects of interest.
 In any case, the use of incorrect SFR(Z,z) clearly affects the resulting cosmological merger rates and may lead to 
 erroneous conclusions. However, the importance of the assumed SFR(Z,z) for calculated $R_{loc}$ was not
 quantified in previous studies.
 \\ 
 Here we focus on the method introduced by \citet{Belczynski16N} 
 (hereafter \citetalias{Belczynski16N}; see appendix on method therein), 
 as it was also used by \citetalias{Chruslinska18} whose models we use in this study.\\
 \citetalias{Belczynski16N} use the cosmic SFR density from \citet{MadauDickinson14} and the mean metallicity 
 of the Universe as found by these authors increased by 0.5 dex to better represent the metallicity at which
 the star formation occurs. 
 This metallicity was used as a mean of the metallicity distribution ($Z_{avg}$), 
 described as a log-normal with a substantial scatter of $\sigma$=0.5 dex.
 Despite the applied shift, the $Z$ of star formation in \citetalias{Belczynski16N} is likely underestimated.
 Observations suggest that massive galaxies dominate the star formation budget in 
 the Universe and the star forming gas found in those
 galaxies has relatively high metal content \citep[which is close to, or higher than the solar value $\Zsun$, 
 even if uncertainty in the absolute metallicity calibration is taken into account, e.g.][]{KewleyEllison08}.
 \\ 
  According to the assumption made by \citetalias{Belczynski16N} $\sim$70\% of stars at certain redshift
   form with log(Z) in the range log(Z$_{avg}$)$\pm$0.5 dex (i.e. range of 1 dex;
 this scatter around log(Z$_{avg}$) is assumed to be constant throughout the cosmic history).
  Using the stellar mass - metallicity relation for galaxies (MZR) by \citet{Tremonti04},
 it can be seen that the range of metallicities corresponding to galaxy stellar masses M$_{\ast}\gtrsim 10^{9}\Msun$ 
 (which are responsible for $\sim$70\% of the local star formation) is around 0.5 dex. 
  \footnote{This range is likely conservative, since MZR found by \citet{Tremonti04} 
  is among the steepest MZRs present in the literature \citep[see][]{KewleyEllison08}.}
 Taking into account the intrinsic scatter in the MZR of $\sim$0.1 dex this range may be broadened to 0.7 dex. 
 This naive estimate suggests that the scatter around Z$_{avg}$ may be smaller than 
  what was assumed in \citetalias{Belczynski16N} (at least in the local Universe, as
 the amount of scatter in the relation may in principle be redshift dependent).
  However, the metallicity gradients within galaxies may also contribute to the scatter in Z at which the stars
  form at a given redshift. Their contribution is difficult to constrain as the results 
  vary significantly between studies and are likely affected by the adopted metallicity calibration and 
  can be mass dependent \citep[e.g.][]{Sanchez-Menguiano16,Belfiore17,Poetrodjojo18}.\\
 In this study we introduce two simple modifications to the assumptions made by \citetalias{Belczynski16N}
to investigate what would be the effect of higher $Z_{avg}$ and smaller scatter around this metallicity on final results.
  
 \section{Method}\label{sec: method}

  \begin{figure}
  \centering
  \includegraphics[scale=0.43]{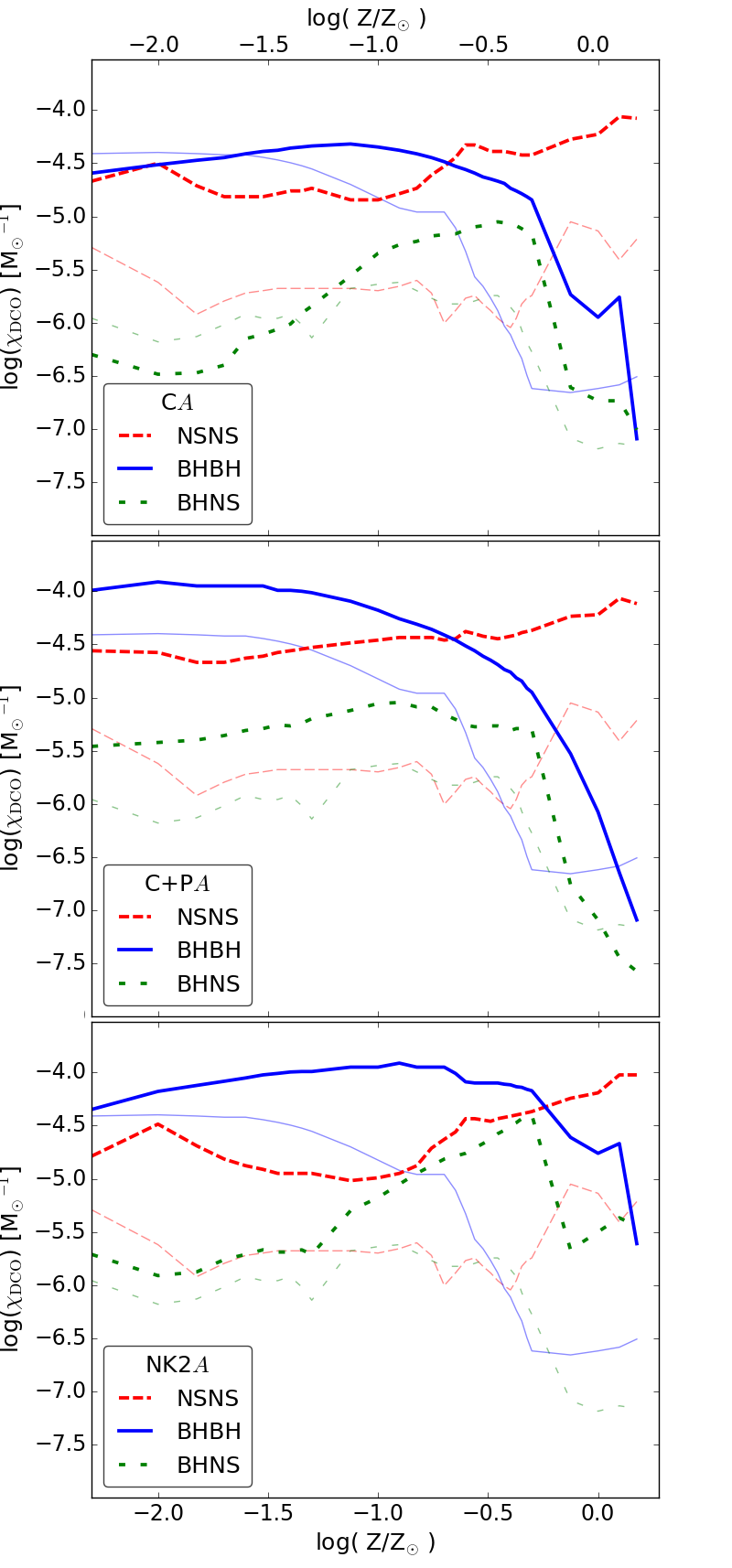}
  \caption{ Formation efficiency $\rm \chi_{DCO}(Z)$  - the number of different types of double compact objects that merge
  within the Hubble time created per unit of mass formed in stars at certain metallicity $Z$ (in solar units, $\Zsun$=0.02) - 
  shown for the three cases (labelled C$A$, C+P$A$ and NK2$A$ as in the original study) identified by \citetalias{Chruslinska18} 
  as producing the highest number of local NSNS mergers within the models probed
  in their study. At the same time, those models were found to overproduce the number of local BHBH mergers.
  For comparison, $\rm \chi_{DCO}(Z)$ for their reference model is shown with the thin line in the background. 
  }
  \label{fig: form_efficiency}
  \end{figure}
  
   \begin{figure}
  \centering
  \includegraphics[scale=0.46]{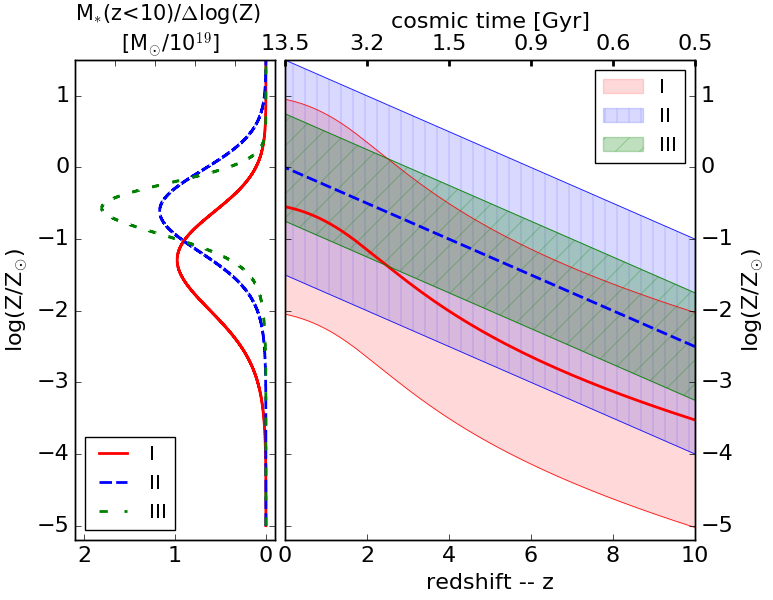}
  \caption{
   The average metallicity Z$_{avg}$ evolution with redshift (thick lines, right panel) 
   for each of the three versions of $\rm f_{sfr}(z,Z)$ distribution describing
   fraction of star formation that at a given $z$ occurs at a certain metallicity (see sec. \ref{sec: method}).
   Version $I$ is identical to the one introduced in \citetalias{Belczynski16N} and used by \citetalias{Chruslinska18}.
   Z$_{avg}$ in cases $II$ and $III$ are identical, but the scatter was reduced twice in $III$.
   The shaded regions indicate 3$\sigma$ spread around Z$_{avg}$.
   The left panel shows the distribution of mass formed in stars since redshift $z$=10 
   at different metallicities for each version of $\rm f_{sfr}(z,Z)$. 
  }
  \label{fig: versions}
  \end{figure}
 
  We take three models (NK2$A$,C$A$,C+P$A$)
  \footnote{$R_{loc}$ calculated for those models can be found in tables 2 and 3 in \citetalias{Chruslinska18}}
  from \citetalias{Chruslinska18} as an example.
  Those models were found to satisfy the current limits on $R_{loc}$ for NSNS systems
  implied by GW170817 \citep[][]{GW170817}, 
  at the same time violating the associated upper limit for $R_{loc}$ of BHBH \citep{GW170104}.
   Briefly, the differences in these models compared to the reference model from \citetalias{Chruslinska18}  are
  \begin{itemize}
  \item in model $NK2$ half of the iron-core collapse supernovae was assumed to lead to small natal kick velocities $\leq 50 km/s$
  \item model $C$ incorporated several modifications found to favour the formation of merging DNS:
   \cite{BrayEldridge16}\footnote{
  Note that after the publication of \citetalias{Chruslinska18},
  \citet{BrayEldridge18} updated the natal kick model given in \cite{BrayEldridge16}. Adopting their updated prescription was 
  found to significantly increase the predicted DNS merger rates \cite{Eldridge18}.
  }
  prescription for the natal kicks that depends on the amount of mass ejected during the supernova and the mass of the remnant
  (as opposed to the distribution proposed by \cite{Hobbs05} used in the reference model, 
  that is independent of the characteristics of the star undergoing supernova), 
  reduced angular momentum loss during the mass transfer and wider limits
  on the helium core mass for the progenitors of stars undergoing electron-capture supernovae;
\item model $C+P$ adds to model $C$ the assumption that mass transfer in systems with Hertzsprung gap donors 
  and NS/BH accretors is stable and never than leads to common envelope. In case of other types of accretors, common
  envelope evolution was allowed.
\end{itemize}
  In all three models the common envelope evolution with Hertzsprung gap donors was allowed 
  (variation $A$, as opposed to variation $B$ from \citetalias{Chruslinska18} where those cases were assumed to lead to merger).
  For more details we refer the reader to the original paper.
  \\
  The formation efficiencies $\rm \chi(Z)$ for those models are shown in fig. \ref{fig: form_efficiency}.
  Note that the simulations were performed for a discrete set of 32 metallicities (listed in \citetalias{Belczynski16N}) 
  and we assume that the formation efficiency within each metallicity bin centred
  at one of these values is the same as for that value. 
  We also assume that $\rm \chi(Z)$ of DCOs at Z$>$Z$_{max}$=0.03 (Z$<$Z$_{min}$=0.0001) are the same as at Z$_{max}$(Z$_{min}$).\\
  As discussed in \ref{sec: rloc and Z}, the mean metallicity of the star formation used in \citetalias{Belczynski16N}
  may lead to an overestimate in the amount of stars forming at low metallicity.
  This effect would be even stronger if the amount of scatter applied to the assumed Z(z) relation proves to be too large.
  Thus, we follow the same procedure as outlined in \citetalias{Belczynski16N} to calculate $R_{loc}$,
  but use three different ways to distribute the cosmic SFR at metallicities, modifying the input Z(z) 
  relation and hence changing $\rm f_{sfr}(z,Z)$: 
  \begin{itemize}
   \item $I$ - identical to the one from \citetalias{Belczynski16N}
   \item $II$ - with higher mean metallicity Z$_{avg}$ (Z$_{avg}\sim \Zsun$ at z=0 
   in contrast to Z$_{avg}\sim 0.3\Zsun$ assumed in \citetalias{Belczynski16N})
   \item $III$ - with Z$_{avg}$ as in $II$ but with twice smaller scatter around the mean
  \end{itemize}
  Those variations are summarised in fig \ref{fig: versions}.
  In version $II$ we use the 'low-end' Z$_{avg}$ introduced by \citet{Dominik13}
  who used the MZR found by \citet{Erb06} and combined it with the average metallicity relation from \citet{Pei99}
  to describe its evolution with redshift. 
  We do not argue that the adopted relation provides the best description of the true metallicity evolution
  of the Universe, but rather use it for its simple form which is sufficient for the purpose of this study.
  The question of distributing the cosmic SFR at different metallicities clearly deserves a more careful investigation. 
  \section{Results}
 \begin{figure*}
\centering
\includegraphics[scale=0.55]{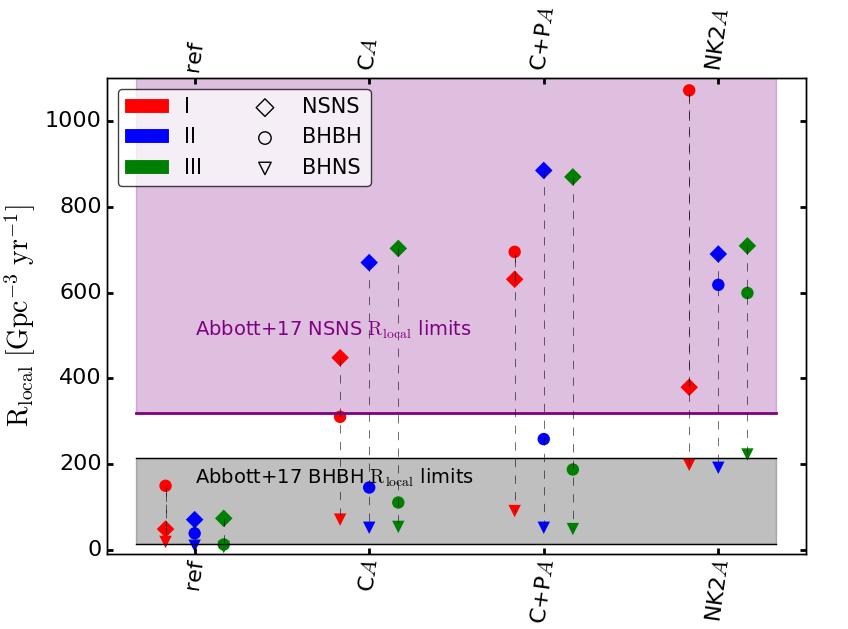}
\caption{ Local merger rate densities ($R_{loc}$) of NSNS (trapezium), BHBH (circle), 
  BHNS/NSBH (triangle) calculated for the three models from \citet{Chruslinska18} that lead to NSNS $R_{loc}$ above 
  the lower limit implied by the gravitational wave observations \citep[purple line;][]{GW170817},
  but the associated BHBH $R_{loc}$ exceed the corresponding upper limit \citep[black line;][]{GW170104}.
  We also show their reference model ($ref$).
  The colours indicate different assumptions about the distribution of the cosmic star formation rate at different metallicities.
  The red points (version I) correspond to the model used in the original study, 
  while the blue ones (II) were calculated assuming higher mean metallicity of the star formation. 
  The green set of results was obtained assuming that the scatter around
  the mean metallicity (described as in II) is twice smaller than in the other two cases.
}
\label{fig: rates}
\end{figure*}

 $R_{loc}$ calculated for the models considered in this study using different $\rm f_{sfr}(z,Z)$ are shown in figure \ref{fig: rates}.
 We also show the reference model (model $ref$ $B$ from \citetalias{Chruslinska18}) for comparison.
 It can be seen that the applied changes in $\rm f_{sfr}(z,Z)$ generally decrease the $R_{loc}$ for BHBH and BHNS binaries,
 while slightly increasing them for NSNS. This is a consequence of both the shape of $\rm \chi_{DCO}(Z)$ 
 and delay time distribution sharply decreasing for long merger times. 
 Shifting Z$_{avg}$(z) to higher values results in smaller population of low-Z binaries that
 contribute to the DCO population that merges locally. 
 Decreasing the width of the metallicity distribution strengthens this effect.
 The NSNS rates in variations $II$ and $III$ increase by a factor of 1.4 - 2 with respect to those calculated in $I$.
 For BHNS binaries this change (decrease) is within a factor of $\sim$3 ($ref$ model), with hardly any difference between the versions in model $NK2A$.
 The BHBH rates decrease by a factor of $\sim$2 (model $NK2A$) up to $\sim$12 ($ref$ model in version $III$).
 Note that these differences are dependent on the $\rm \chi_{DCO}(Z)$ that results from population synthesis calculations and 
 the examples presented in this study sample only a small fraction of the parameter space involved in such calculations.

\begin{figure}
\centering
\includegraphics[scale=0.5]{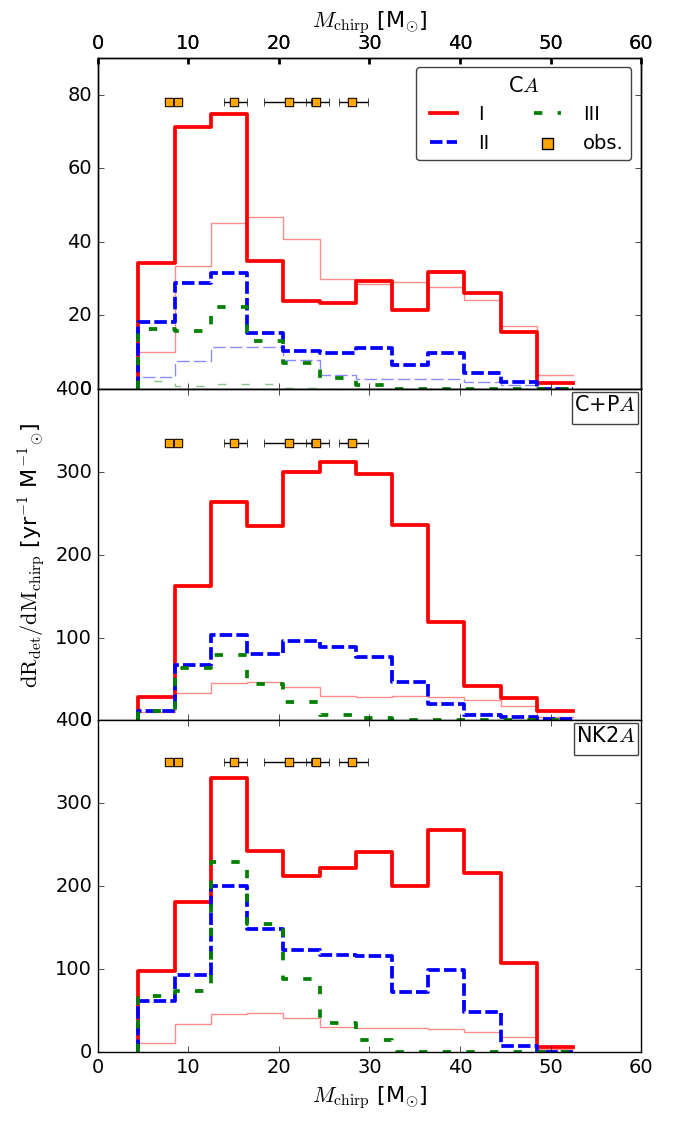}
\caption{
The detection rate R$_{det}$ weighted chirp-mass distribution of the BHBH binaries merging locally
for the models considered in this study (different panels) and for the three versions
of the distribution of the cosmic star formation rate at different metallicities
(different colours).
The orange squares mark chirp masses of BHBH mergers observed in gravitational waves so far
\citep[and a candidate detection LVT151012 at $M_{chirp}\sim$15$\Msun$;][]{GWO1,GW170104,GW170814,GW170608}.
R$_{det}$ were calculated assuming detection distances for NS-NS mergers of 170 Mpc.
The reference model is plotted in the background (thin lines).
}
\label{fig: BHBH_Mchirp}
\end{figure}

 In variation $III$ for models $CA$ and $C+PA$ the rates for all DCO types are consistent with gravitational waves limits
 \footnote{12-213 $\gpy$ for BHBH \citep{GW170104}, 320 - 4740 $\gpy$ for NSNS \citep{GW170817} and an upper limit of 3600 $\gpy$ 
 for BHNS/NSBH \citep{Abbott16_limits}}.
 This is also true for model $CA$ in variation $II$, where only the mean Z$_{avg}$ was increased.
 For the model $NK2A$ the formation efficiency of merging BHBH remains high up to solar-like metallicities.
 In this model NSNS and BHBH $R_{loc}$ observational limits likely cannot be met simultaneously for 
 the same set of evolution-related assumptions for any reasonable model of the SFR history and chemical evolution of the Universe.
  \\ \newline
  In general, mergers of more massive binaries can be detected from larger distances and hence 
  $R_{loc}$ does not translate directly to the observed frequency of mergers (detection rate),
  which scales with a combination of masses of merging objects.
  This effect is mostly important for BHBH binaries that can form with a wide range of masses.
  Since the most massive black holes are expected to form at low metallicities, 
  modifications in $\rm f_{sfr}(z,Z)$ have important consequences for the mass 
  (either total or chirp mass $\mathcal{M}_{chirp}$) distribution of merging BHBH binaries. 
  This distribution extends to higher masses if more (recent) SFR happens at low metallicities 
  and hence the average mass in variations $II$ and $III$ is lower than in $I$.
  As a consequence, the decrease in the detection rates for BHBH estimated for $II$ or $III$ with respect to $I$
  would be bigger than in their $R_{loc}$.  
  \\
  The $\mathcal{M}_{chirp}$ distribution of merging binaries can be probed with gravitational wave observations
  which provides additional constraint on our models. Thus, we need to verify if the discussed models agree with the
  $\mathcal{M}_{chirp}$ distribution of BHBH mergers detected so far.
  In figure \ref{fig: BHBH_Mchirp} we show the detection rate-weighted distribution of BHBH $\mathcal{M}_{chirp}$ 
  for different models and $\rm f_{sfr}(z,Z)$ variations.
  The approximate detection rates were calculated using eq. 5 from \citetalias{Chruslinska18} \citep[see sec. III in][]{Abadie10}
  assuming double neutron star detection range of 170 Mpc.
  It can be seen that in version $III$ the detection of BHBH merger with $\mathcal{M}_{chirp}\gtrsim$30$\Msun$ is unlikely
  in all of the models considered in this study, while for $I$ and $II$ the distribution extends up to 
  $\mathcal{M}_{chirp}\sim$50$\Msun$  
  \footnote{the effect of pair instability mass loss was not included in the models presented in this study, however it affects only
   the most massive BHs (M$\gtrsim 40\Msun$) forming at very low metallicities and their contribution to the population merging
   locally is negligible \citep[see][]{Belczynski16PISN}}.
  While those distributions can be probed by the future observations, for now the observed sample
  is too small to allow for any firm conclusions from the comparison.
  For instance, two sample KS test performed on each of the model distributions shown in fig. \ref{fig: BHBH_Mchirp}
  and the observed sample does not allow to rule out any of those distributions at the confidence level
  higher than 96.4\%, with the lowest p-values revealed by models $NK2A$ $I$ ($\sim$0.036) and $C+PA$ in $I$ ($\sim$0.056).

 \section{Conclusions}
 
  Using three models from \citetalias{Chruslinska18} as an example,
  we have demonstrated the importance of the assumptions related to the star formation history and chemical evolution
  of the Universe for the obtained properties of merging populations of double compact objects,
  in particular for the estimated merger rate densities.
  Those models were found to lead to the local NSNS merger rate density consistent with the current limits 
  from gravitational wave observations, at the same time overproducing the number of the local BHBH mergers.
  One possible solution to this conundrum, as suggested by \citetalias{Chruslinska18}, 
  is that BH form with higher natal kicks than assumed in those models, or the common envelope evolution is different for
  massive BHBH progenitors than for NSNS progenitors. \\
  Differences in $\rm f_{sfr}(z,Z)$ (or more generally in SFR(Z,z)) induce differences in the properties of the population of 
  merging DCOs observed at a certain redshift. Since the formation efficiency of merging DCOs behaves differently
  with changing metallicity for different types of systems, the
  change in SFR(Z,z) affects the ratios of numbers of DCOs of different types and hence their merger rates.
  The $\rm f_{sfr}(z,Z)$ assumed in the original study likely overestimates the amount of stars forming at low metallicity.\\
  We have shown that when the average metallicity of the star formation is increased to the values more consistent with
  observations of local galaxies, the number density of local BHBH mergers decreases sufficiently to match the observational limits
  in one of the models ($CA$).
  In section \ref{sec: rloc and Z} we argued that the amount of scatter around the average used in the original method
  may be overestimated if metallicity gradients within the regions responsible for the bulk of SFR in the galaxies
  are sufficiently small.  
  If this scatter is reduced, R$_{loc}$ in the model $C+PA$ also meet the gravitational
  wave limits. 
  In the remaining case the observed R$_{loc}$ likely cannot be reproduced simultaneously for all types of DCO by the use of any 
  reasonable SFR(Z,z) distribution.\\
  Note that the formation efficiency of DCOs is model dependent and so is the change in R$_{loc}$ in response to change in $\rm f_{sfr}(z,Z)$.
  The models used in this study sample only a small part of the parameter space involved in population synthesis calculations.
  We do not argue that they provide the correct description of the DCO population, but rather use them as a good example showing 
  how the adopted assumptions about $\rm f_{sfr}(z,Z)$ add to degeneracies in the conclusions and final results of those calculations.
  \\
  Changes in SFR(Z,z) also have important consequences for the mass (chirp mass) distribution of merging BHBH,
  which will be sampled with the gravitational wave observations in the future.
  \\
  Our findings highlight the importance of the choice of a particular way to distribute the cosmic star formation rate across 
  metallicities and time and the need to better understand the uncertainties associated with that choice. 
  Without tighter constraints on this distribution one has to deal with another layer of degeneracy e.g. 
  in the calculated merger rates,
  besides degeneracies connected to the description of various evolutionary phases of DCO progenitors,
  which hinders drawing any strong conclusions from studies that aim to use cosmological rates as constraints.
 
\section*{Acknowledgements}
MC and GN acknowledge support from the Netherlands Organisation for Scientific Research (NWO).
KB acknowledges support from the Polish National Science Center (NCN) grants
Sonata Bis 2 (DEC-2012/07/E/ST9/01360), OPUS (2015/19/B/ST9/01099), Maestro 
2015/18/A/ST9/00746 and LOFT/eXTP 2013/10/M/ST9/00729.


\bibliographystyle{mnras}
\bibliography{bibliography} 


%


\bsp	
\label{lastpage}
\end{document}